\documentstyle{article}
\font\tenrm=cmr10
\font\tenit=cmti10
\font\elevenbf=cmbx10 scaled\magstep 1
\font\elevenrm=cmr10 scaled\magstep 1
 1

\def\Re{{\cal R \mskip-4mu \lower.1ex \hbox{\it e}\,}}
\def\Im{{\cal I \mskip-5mu \lower.1ex \hbox{\it m}\,}}
\def\ie{{\it i.e.}}

\def\sub#1{_{\lower.25ex\hbox{$\scriptstyle#1$}}}
\def\sul#1{_{\kern-.1em#1}}
\def\sll#1{_{\kern-.2em#1}}
\def\sbl#1{_{\kern-.1em\lower.25ex\hbox{$\scriptstyle#1$}}}
\def\ssb#1{_{\lower.25ex\hbox{$\scriptscriptstyle#1$}}}
\def\sbb#1{_{\lower.4ex\hbox{$\scriptstyle#1$}}}

\def\gev{\,{\rm GeV}}

\def\to{\rightarrow}
\def\mh{\ifmmode m\sbl H \else $m\sbl H$\fi}
\def\mch{\ifmmode m_{H^\pm} \else $m_{H^\pm}$\fi}
\def\mt{\ifmmode m_t\else $m_t$\fi}
\def\mc{\ifmmode m_c\else $m_c$\fi}
\def\mz{\ifmmode M_Z\else $M_Z$\fi}
\def\mw{\ifmmode M_W\else $M_W$\fi}
\def\mws{\ifmmode M_W^2 \else $M_W^2$\fi}
\def\mhs{\ifmmode m_H^2 \else $m_H^2$\fi}
\def\mzs{\ifmmode M_Z^2 \else $M_Z^2$\fi}
\def\mts{\ifmmode m_t^2 \else $m_t^2$\fi}
\def\mcs{\ifmmode m_c^2 \else $m_c^2$\fi}
\def\mchs{\ifmmode m_{H^\pm}^2 \else $m_{H^\pm}^2$\fi}
\def\ztwo{\ifmmode Z_2\else $Z_2$\fi}
\def\zone{\ifmmode Z_1\else $Z_1$\fi}
\def\mtwo{\ifmmode M_2\else $M_2$\fi}
\def\mone{\ifmmode M_1\else $M_1$\fi}
\def\tb{\ifmmode \tan\beta \else $\tan\beta$\fi}
\def\xw{\ifmmode x\sub w\else $x\sub w$\fi}
\def\ch{\ifmmode H^\pm \else $H^\pm$\fi}
\def\lum{\ifmmode {\cal L}\else ${\cal L}$\fi}
\def\inpb{\ifmmode {\rm pb}^{-1}\else ${\rm pb}^{-1}$\fi}
\def\infb{\ifmmode {\rm fb}^{-1}\else ${\rm fb}^{-1}$\fi}
\def\epem{\ifmmode e^+e^-\else $e^+e^-$\fi}
\def\ppb{\ifmmode \bar pp\else $\bar pp$\fi}

\newskip\zatskip \zatskip=0pt plus0pt minus0pt
\def\matth{\mathsurround=0pt}
\def\lsim{\mathrel{\mathpalette\atversim<}}
\def\gsim{\mathrel{\mathpalette\atversim>}}
\def\atversim#1#2{\lower0.7ex\vbox{\baselineskip\zatskip\lineskip\zatskip
  \lineskiplimit 0pt\ialign{$\matth#1\hfil##\hfil$\crcr#2\crcr\sim\crcr}}}

\renewenvironment{thebibliography}[1]
 { \elevenrm
   \begin{list}{\arabic{enumi}.}
    {\usecounter{enumi} \setlength{\parsep}{0pt}
     \setlength{\itemsep}{3pt} \settowidth{\labelwidth}{#1.}
     \sloppy
    }}{\end{list}}

\renewcommand{\thefootnote}{\fnsymbol{footnote}}

\begin{document} \begin{titlepage}
\rightline{\vbox{\halign{&#\hfil\cr
&ANL-HEP-CP-92-126\cr
&November 1992\cr}}}
\vspace{1in}
\begin{center}

{\Large\bf  SIGNALS FOR VIRTUAL LEPTOQUARK
               EXCHANGE AT COLLIDERS }\footnote{Research supported by
the U.S. Department of
Energy, Division of High Energy Physics, Contract W-31-109-ENG-38.}

\bigskip

\normalsize
M.A. DONCHESKI\\

\smallskip
Department of Physics\\
University of Wisconsin\\
Madison, WI 53706 USA\\
\smallskip

and
\smallskip

J.L. HEWETT\\

\smallskip
High Energy Physics Division\\
Argonne National Laboratory\\
Argonne, IL 60439\\

\end{center}
\bigskip

\begin{abstract}

We study the effects of virtual leptoquark exchange on charged current and
neutral current processes at HERA, on di-lepton production at the Tevatron, and
on quark pair production at LEP II.  We present the areas of
parameter space that can be excluded at these colliders
by searching for deviations
from Standard Model expectations.

\end{abstract}

\vskip1.75in

\noindent{Presented by J.L.H. at the
{\it 1992 Meeting of the Division of Particles
and Fields},
Fermilab, Batavia, IL, November 10-14, 1992. }

\renewcommand{\thefootnote}{\arabic{footnote}} \end{titlepage}

\begin{center}{{\elevenbf SIGNALS FOR VIRTUAL LEPTOQUARK
               EXCHANGE AT COLLIDERS \\}
\vglue 0.6cm
{\tenrm M.A. DONCHESKI$^a$ and J.L. HEWETT$^b$ \\}
\baselineskip=13pt
{\tenit $^a$Department of Physics, University of Wisconsin, Madison,
WI 53706 USA\\}
\baselineskip=12pt
{\tenit $^b$High Energy Physics Division, Argonne National Laboratory \\}
\baselineskip=12pt
{\tenit 9700 S.\ Cass Ave., Argonne, IL  60439  USA\\}
\vglue 0.2cm
{\tenrm ABSTRACT}}
\end{center}
\vglue 0.1cm
{\rightskip=3pc
 \leftskip=3pc
 \tenrm\baselineskip=12pt
 \noindent
We study the effects of virtual leptoquark exchange on charged current and
neutral current processes at HERA, on di-lepton production at the Tevatron, and
on quark pair production at LEP II.  We present the areas of
parameter space that can be excluded at these colliders
by searching for deviations
from Standard Model expectations.
\vglue 0.6cm}
\elevenrm
Many theories which go beyond the Standard Model (SM) are inspired by the
symmetry between the quark and lepton generations and try to relate them at a
more fundamental level.  As a result, many of these models contain new
particles, called leptoquarks, which naturally couple to a lepton-quark pair.
These particles need not
be heavy; in fact, leptoquarks can have a mass $\sim 100\gev$ and still
avoid\cite{buch} conflicts with rapid proton decay and dangerously large
flavor changing neutral currents.  This is particularly true in models where
each generation of fermions has its own leptoquark(s) which couples only
within that generation.
Here, results are presented for the leptoquark present in
superstring-inspired $E_6$ models\cite{esix}.  These leptoquarks,
denoted by $S$, are scalar, charge $-1/3$, baryon number $=+1/3$, lepton
number $=+1$, weak iso-singlets.
Their interactions are governed by the $E_6$ superpotential terms
\begin{equation}
\lambda_LLS^cQ + \lambda_RSu^ce^c + \lambda'\nu^cSd^c \,,
\end{equation}
where $L$ and $Q$ represent the left-handed lepton and quark doublets,
respectively, and the superscript $c$ denotes the charge conjugate states.
The Yukawa couplings, \ie, the $\lambda$'s, are {\it a priori} unknown.
For calculational purposes, the Yukawa couplings are parameterized by
(with $F_{L}=F_{R}\le 1$)
\begin{equation}
{\lambda_{L,R}^2\over 4\pi} = F_{L,R}~~\alpha \,.
\end{equation}

At \epem\ colliders, direct leptoquark pair production can proceed through
s-channel $Z$ and $\gamma$ exchange.  The search region is essentially
set by the kinematic reach of the collider, \ie, $M_S\lsim\sqrt s/2$.
However, if a leptoquark is too heavy to be produced directly, perhaps
it can be detected through its indirect effects via virtual exchange.
This is similar in nature to the detection of the
SM $Z$ boson at PEP/PETRA.  In \epem\ collisions, leptoquarks
can contribute to the process $\epem\to q\bar q$ through t-channel exchange
via their Yukawa couplings.  The $95\%$ C.L. bounds which can be
obtained\cite{tgr}
at LEP II from such reactions is presented in Fig.\ 1 in the leptoquark
coupling strength - mass plane for various values of integrated luminosity.
Here, the search region corresponds to the area above the curves.

Direct leptoquark production at hadron colliders proceeds through $gg$ fusion
and $q\bar q$ annihilation and is also independent of the unknown Yukawa
couplings.  Present searches at the Tevatron have placed\cite{teva} the  bound
$M_S > 113\gev$.  Leptoquarks can also contribute to the reverse of the process
discussed above, \ie, $q\bar q\to\epem$.  However, preliminary investigation
shows\cite{mad} that it is difficult to distinguish the leptoquark signal
from the background in this case.

By its nature, the high-energy ep collider HERA is especially well-suited to
study leptoquarks.  Direct production can occur through an $s$-channel
resonance via the Yukawa couplings at enormous rates with distinctive peaks
in the $x$-distribution.  HERA experimental searches are
expected\cite{hera} to reach a discovery limit
of $\sim 250\gev$ when the leptoquark
coupling strength is of order $0.01\alpha_{em}$, and up to the kinematic limit
if the coupling is equal to $\alpha_{em}$.
If leptoquarks are too massive to
be produced directly at HERA, they again can be detected through their
indirect effects via virtual exchange by searching for deviations from SM
expectations for certain processes.  Several
authors\cite{virt} have examined such effects on the neutral current
asymmetries that can be formed with polarized electron beams, and have found
that departures from the SM are small, even for leptoquarks of low mass
({\it e.g.}, $\sim 400\gev$) and large couplings.  The possible indirect
effects on charged current as well as neutral current processes have also
been systematically investigated\cite{jlh}.  The best results are obtained by
examining the ratio $R=\sigma_{NC}/\sigma_{CC}$, where discovery limits can
reach leptoquark masses of order $800\gev$ for electromagnetic coupling
strengths (with $200\inpb$ of integrated luminosity per $e^+, e^-$ beam).
Taking the ratio of neutral to charged current cross sections is also
advantageous because several systematic uncertainties, as well as those from a
lack of detailed knowledge of the parton distributions, will cancel.
Neutral current events, \ie, the reactions
$e^\pm  q\to e^\pm  q$ and $e^\pm\bar q\to e^\pm\bar q$, occur through
$t$-channel $\gamma$ and $Z$ exchange in the SM.  Leptoquarks can also
contribute via
$s$- and $u$-channel exchanges.  In the charged current case,
\ie, $e^\pm q\to \nu q$ and $e^\pm \bar q\to \nu\bar q$,  leptoquarks can
also contribute in these same channels.
In order to ensure that the NC and CC events are cleanly separated and
identified,
cuts are imposed on the scaling variables $x$ and $y$, as well as on
the  transverse momentum of the out-going lepton, $p_T(e)>5\gev $ and
$\not p_T(\nu)>20\gev$.  The
$90\%$ and $95\%$ C.L.  discovery
region in the leptoquark coupling-mass plane, based on measurements of this
observable is displayed in Fig. 2.
Figure 2 shows the results
for unpolarized $e^\pm$ beams with integrated luminosities of $20,~200,$ and
$500\inpb$ per beam.
The region that can be explored with $200\inpb$
per unpolarized beam is $m_{S} \lsim 800\gev$ for large
leptoquark-electron-quark couplings ($F \sim  1$), and $F \gsim 0.13$ for
$m_{S} \sim 314\gev$.

In summary, we find that virtual leptoquark exchange can have significant
contributions to processes in \epem\ and $ep$ colliders.  We urge
our experimental colleagues to search for these effects! \\
%
\def\MPL #1 #2 #3 {Mod.~Phys.~Lett.~{\bf#1},\ #2 (#3)}
\def\NPB #1 #2 #3 {Nucl.~Phys.~{\bf#1},\ #2 (#3)}
\def\PLB #1 #2 #3 {Phys.~Lett.~{\bf#1},\ #2 (#3)}
\def\PR #1 #2 #3 {Phys.~Rep.~{\bf#1},\ #2 (#3)}
\def\PRD #1 #2 #3 {Phys.~Rev.~{\bf#1},\ #2 (#3)}
\def\PRL #1 #2 #3 {Phys.~Rev.~Lett.~{\bf#1},\ #2 (#3)}
\def\RMP #1 #2 #3 {Rev.~Mod.~Phys.~{\bf#1},\ #2 (#3)}
\def\ZPC #1 #2 #3 {Z.~Phys.~{\bf#1},\ #2 (#3)}
\def\IJMP #1 #2 #3 {Int.~J.~Mod.~Phys.~{\bf#1},\ #2 (#3)}
%
\\
{\elevenbf\noindent  References \hfil}
\vglue 0.2cm

\newpage

\noindent
{Fig. 1. $95\%$ C.L. discovery region at LEP II (with $\sqrt s=200\gev$)
in the leptoquark coupling - mass plane for various values of integrated
luminosity.  Here, $\kappa=2F$ in the notation presented in the text.}

\noindent
{Fig. 2.  Discovery region in the $m_{S}-F$ plane at HERA from the
ratio, $R$, for
different integrated luminosities as shown.  The area to the upper left of the
solid (dashed) curves can be excluded at the 90\% (95\%) C.L.}

\end{document}